\documentclass[twocolumn,showpacs,aps]{revtex4}
\usepackage{epsfig}

\begin{document}

\title{Gate capacitance of back-gated nanowire field-effect
  transistors}

\author{Olaf Wunnicke}

\affiliation{Philips Research Laboratories, High Tech Campus 4, 5656
AE Eindhoven, The Netherlands}

\begin{abstract}
Gate capacitances of back-gated nanowire field-effect transistors
(NW-FETs) are calculated by means of finite element methods and the
results are compared with analytical results of the ``metallic
cylinder on an infinite metal plate model''. Completely embedded
and non-embedded NW-FETs are considered. It is shown
that the use of the analytical expressions also for
non-embedded NW-FETs gives carrier mobilities that are nearly two
times too small. 
Furthermore, the electric field amplification of non-embedded
NW-FETs and the influence of the cross-section shape of the nanowires
are discussed. 
\end{abstract}

%
%


\maketitle

Semiconducting nanowires (NWs) with diameters from several nanometers
up to around 100nm are currently being investigating  for their
potential towards realization of electronic devices with significantly
enhanced 
performance like enhanced carrier mobility. While much effort is spent
on growth and structural characterization, obviously reliable and
accurate electrical characterization is equally important. For
electronic applications the charge carrier mobility $\mu$ is a crucial
property. The mobility can be determined from the linear part of the
transconductance  
$g = dI_{sd}/dU_{gs}$ measured in a field-effect transistor device
$\mu = g L^2 / (C U_{sd}),$
with $L$ being the channel length, $C$ the gate capacitance,
$U_{sd}$ the source-drain, and $U_{gs}$ the gate-source voltage. 

\begin{figure}[h!]
  \begin{center}
    \epsfig{file=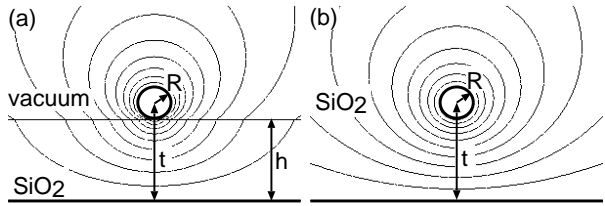,width=8cm,angle=0}
    \caption{Cross section geometry of back-gated NW-FETs of
    non-embedded NWs (a) and embedded NWs (b) with 
    $t/R = 6$.  The lines are on equally separated constant potentials
    obtained by FEM calculations.} 
    \label{fig.SiO2_equipot}
  \end{center}
\end{figure}

In the case of NWs usually back-gated 
nanowire field-effect transistors (NW-FETs) are fabricated:
\cite{FD03,TD97,BT04,LX05,WW03,Av02,KP03,LL04,PK04,CD00,MS98,LY05,CZ03}
a highly doped Si-substrate and thermally grown SiO$_2$ function as
the back-gate and the gate dielectric, respectively. Then NWs
are transferred on the SiO$_2$ followed by the deposition of the
source and drain electrodes. 
Such a transistor has the advantage that it can be processed
relatively easily. However, in practical devices other geometries like
the wrap-around NW-FET will be  preferred. 
A cross sectional view along the NW axis with equipotential lines due
to the gate voltage is 
shown in Fig. \ref{fig.SiO2_equipot}(a).

To calculate the gate capacitance, usually the ``metallic cylinder on
an infinite metal plate model'' \cite{RW93} is used.
\cite{LX05,WW03,Av02,KP03,LL04,PK04,CD00,MS98,LY05,CZ03} 
The cross section geometry and the equipotential lines of this model
are depicted in Fig. \ref{fig.SiO2_equipot}(b).
To use this model, the charge density in the nanowire is assumed to be
so high that the semiconducting NW can be treated as metallic.  
It was shown \cite{VS06} that this approximation yields reasonable
results for GaN NWs with doping concentrations above
$10^{17}\mathrm{cm}^{-3}$. 
Additionally, it is assumed that the NW is much longer than the
dielectric layer thickness so that the fringing capacitance at the
source and drain contact can be neglected and that there are no
movable charges or defects in the dielectric layer or at the NW
surface. The model further assumes that the NWs are 
completely embedded in the dielectric and posses a circular cross
section. The latter assumptions are reexamined in detail in this
paper.

\begin{figure}[h!]
  \begin{center}
    \epsfig{file=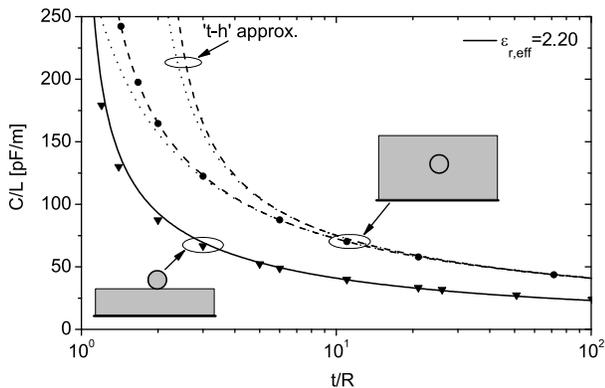,width=8cm,angle=0}
    \caption{Gate capacitances with SiO$_2$ gate dielectric. 
    Analytical results for embedded NW-FETs are shown by
    a dashed line [Eq. (\ref{eq.acosh})] and by a dotted line 
    [Eq. (\ref{eq.ln})].
    FEM calculations of embedded and non-embedded NWs are represented
    by circles and triangles, respectively.
    The solid line is a fit of the numerical results. 
    ``t-h approx'' marks analytical results with $t$ approximated by $h$ in
    Eq. (\ref{eq.acosh}) and Eq. (\ref{eq.ln}).} 
    \label{fig.SiO2}
  \end{center}
\end{figure}

The model yields an analytical equation for the gate capacitance per
unit length
\begin{equation}
\frac{C}{L} = \frac{2\pi \epsilon_0
\epsilon_r}{\cosh^{-1} \left(\frac{t}{R}\right)},
\label{eq.acosh}
\end{equation}
with $\epsilon_r$ being the dielectric constant of the embedding
dielectric, $t$ the distance between the metal plate and the center
of the cylinder, and $R$ the radius of the cylinder [see
Fig. \ref{fig.SiO2_equipot}(b)].
For $x= t/R \gg 1$, the approximation 
$\cosh^{-1} (x) = \ln\left(x+\sqrt{x^2-1}\right) \approx \ln (2x)$ 
can be used. Equation (\ref{eq.acosh}) then reads
\cite{MF53}
\begin{equation}
\frac{C}{L} = \frac{2\pi \epsilon_0
\epsilon_r}{\ln\left(2 \frac{t}{R}\right)}.
\label{eq.ln}
\end{equation}
In Fig. \ref{fig.SiO2} the results of Eqs. (\ref{eq.acosh})
and (\ref{eq.ln}) are shown.
Here SiO$_2$ is considered as the gate dielectric with $\epsilon_r =
3.9$. The relative error by using Eq. (\ref{eq.ln}) is less than 1\% for
$t/R > 6$. Because most NW-FETs have $t/R$ ratios above six, this
approximation yields usually good accuracy.

Equation (\ref{eq.acosh}) \cite{LX05,WW03} and Eq. (\ref{eq.ln})
\cite{Av02,KP03,LL04,PK04,CD00,MS98,LY05,CZ03} are used to estimate
the gate capacitance also in the case of back-gated NW-FETs in which
the NW is lying on top of the gate dielectric 
[Fig. \ref{fig.SiO2_equipot}(a)]. Because of the missing dielectric
material around the NW, it is expected that both analytical equations
give only an upper limit for the gate capacitance. 
To examine this in detail, I have carried out finite element method
(FEM) studies \cite{FEM} of the gate capacitance.

The FEM results are shown in Fig. \ref{fig.SiO2}.
For the embedded NWs the numerical results agree very well
with the analytical ones using Eq. (\ref{eq.acosh}). For small $t/R$
values the deviation due to the approximation in (\ref{eq.ln}) can be
seen. For the non-embedded NWs, the capacitances are nearly
two times lower than the analytical results for embedded
NWs. Thus if Eqs. (\ref{eq.acosh}) or (\ref{eq.ln}) are used to
determine the gate capacitance also in the case of non-embedded 
NW-FETs, the carrier mobilities 
are underestimated by a factor of almost two.
This discrepancy is very important when the carrier mobilities in
NWs are compared with bulk values. Furthermore, this implies
that the gate coupling of non-embedded NW-FETs can be nearly doubled
by packing the NWs completely in SiO$_2$. This has also a
positive effect on the electric field distribution as will be
discussed later. 

In literature the weaker gate coupling in case of non-embedded NW-FETs
is taken into account by an empirical effective dielectric constant
for SiO$_2$ of
$\epsilon_{r,\mathrm{eff}}=2.5$ \cite{MS98,LY05} or
$\epsilon_{r,\mathrm{eff}}=1.95$ \cite{VS06} in combination with
Eq. (\ref{eq.ln}). In Fig. \ref{fig.SiO2}, a fit of the numerical data
using an effective dielectric constant is shown.
The fit is carried out for $6 < t/R < 100$ and results in
$\epsilon_{r,\mathrm{eff}}=2.20$ with a relative deviation of less
than 3\% in this $t/R$ range. The 
reason why the use of an effective dielectric constant works well, is
the relatively small difference between the dielectric constants of
SiO$_2$ and vacuum. This ensures that the $t/R$ dependence of the gate
capacitance is not changed much and only the gate coupling is reduced.

\begin{figure}[h!]
  \begin{center}
    \epsfig{file=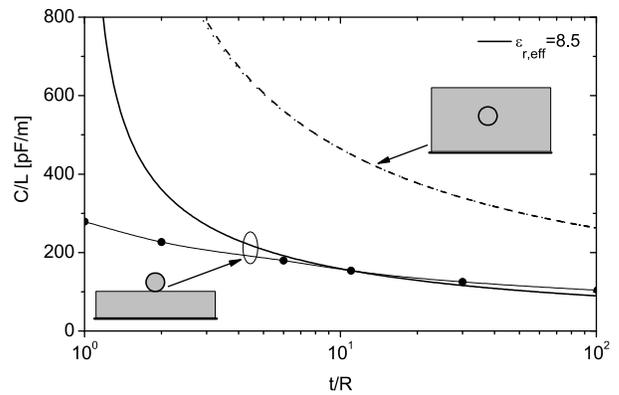,width=8cm,angle=0}
    \caption{Gate capacitances with HfO$_2$ gate dielectric.
    Analytical results of Eq. (\ref{eq.acosh}) (dashed line) and
    Eq. (\ref{eq.ln}) (dotted line) lie nearly on top of each other. 
    FEM calculations of non-embedded NWs are represented by circles.
    The thick solid line is a fit of the numerical results.
    The thin line is a guide to the eyes.} 
    \label{fig.HfO2}
  \end{center}
\end{figure}

For comparison, gate capacitances for the high-$k$ dielectric
HfO$_2$ ($\epsilon_r=25$) are calculated (Fig. \ref{fig.HfO2}).  
The fit of the calculated capacitances in the range $6 < t/R < 100$
results in an effective dielectric constant of
$\epsilon_{r,\mathrm{eff}} = 8.5$.
But clearly this fit does not
describe the gate capacitances well. The reason is
that the electric field is changed drastically compared to the
embedded case due to the high dielectric constant. For high-$k$
dielectrics, the radius of the NW becomes 
less important for the gate capacitance since the electric field gets
more and more confined to the gate dielectric. 
This results in a weaker $t/R$ dependence
compared with SiO$_2$.

A further point of interest is that much higher electric fields are
reached when the NW is not embedded in the gate dielectric. 
The highest electric fields are near the contact between the NW
and the gate dielectric. 
Assuming $t=120$nm and $R=20$nm as in Fig. \ref{fig.SiO2_equipot}, 
the dielectric strength of SiO$_2$ ($10^7$V/cm) is reached
locally at $U_{gs}=40$V for the embedded NW, but
already at $U_{gs}=6$V for the non-embedded NW. For comparison, a
100nm thick SiO$_2$ plate capacitor withstands roughly 100V. This is
an important issue, especially if the gate hysteresis \cite{FK02} and
the durability are considered. 

\begin{figure}[h!]
  \begin{center}
    \epsfig{file=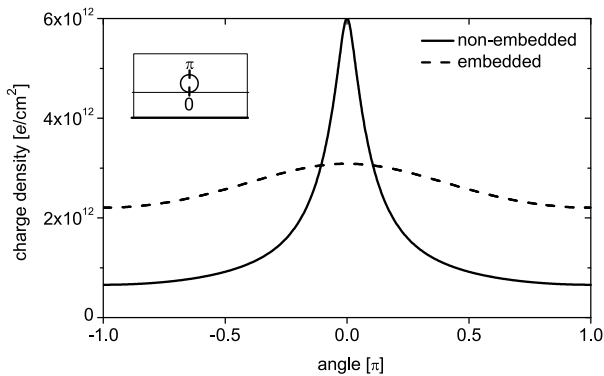,width=8cm,angle=0}
    \caption{FEM calculations of the induced charge density in
    elementary charges per cm$^2$ on the NW
    surface for non-embedded (solid line) and embedded (dashed line)
    NW-FETs with $R=20$nm, $t=120$nm, and $U_{gs}=6$V.} 
    \label{fig.surf_E}
  \end{center}
\end{figure}

To obtain a high on-current in the NW-FET, a homogeneously induced
charge density on the NW surface is important. The optimum case is
obtained with a wrap-around NW-FET. 
In Fig. \ref{fig.surf_E}, the induced charge densities are shown
for $U_{gs} = 6$V. 
The charge density of the non-embedded NW-FET possess the same trend
as the embedded one  
only at a lower level due to the weaker gate coupling. But at the
contact point, the non-embedded NW shows an even stronger peak than
the embedded NW because of the high electric fields.
Importantly, the maximum amount of charges that can be induced in the
embedded NW-FET is around 12 times higher than in non-embedded one for
the above geometry. This is due  
to the higher gate capacitance and due to the higher maximum gate
voltage. This voltage is scalable, if the $t/R$ ratio is kept
constant. 

In addition, in various instances \cite{LX05,LL04,PK04,CD00,MS98,LY05}
the distance between the 
center of the NW and the back gate $t$ has moreover been 
approximated by the thickness of the gate dielectric $h = t - R$.
In Fig. \ref{fig.SiO2}, the analytical values of Eqs. 
(\ref{eq.acosh}) and (\ref{eq.ln}) while using this approximation are
shown by ``t-h approx''. For $t/R > 10$, the error is negligible but
it becomes large for smaller values. This can result in values for the
gate capacitance that are off by more than a factor of three.

Usually NWs grow with a non-cylindrical cross section. 
The actual shape rather depends on the crystal structure and the
growth direction. For zinc-blende NWs grown in the [111] direction, a
hexagonal cross section is found while NWs grown in the [001]
direction have a square-like cross section. Furthermore, NWs grown in
the [112] direction with a triangular cross section have been
observed. \cite{VI06} 
\begin{figure}[h!]
  \begin{center}
    \epsfig{file=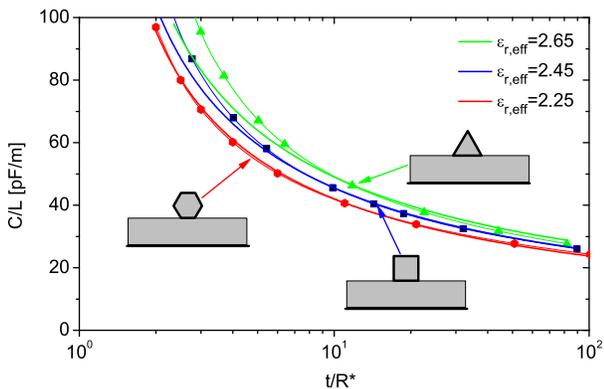,width=8cm,angle=0}
    \caption{FEM calculations of gate capacitances with SiO$_2$ of
    non-embedded NWs with hexagonal (red), square-like (blue),
    and triangular (green) cross sections (from bottom to top). The
    thick lines are fits of the numerical data. The thin lines are
    guides to the eyes. $R^\star$ is the effective radius of a circle
    with the same area as the hexagon, the square, or the triangle,
    respectively. (color online)}
    \label{fig.SiO2_growth}
  \end{center}
\end{figure}
In Fig. \ref{fig.SiO2_growth}, the calculated gate capacitances for
NWs with these cross sections are shown.
To make the results comparable, I define an effective radius $R^\star$
as the radius of a circle with the same area as the hexagon, the
square, or the triangle. The fits are carried out for $6 < t/R^\star <
100$ and result in an effective 
dielectric constant of $\epsilon_{r,\mathrm{eff}}=2.25$ for the
hexagonal, $\epsilon_{r,\mathrm{eff}}=2.45$ for the square-like, and
$\epsilon_{r,\mathrm{eff}}=2.65$ for the triangular cross section.
The higher effective dielectric constants are due to the larger
contact area between these NWs and the gate dielectric compared
to the line contact for NWs with circular cross section. This
effect increases for smaller $t/R^\star$ ratios.
Because of the sharp corners, the electric field amplification is an
important issue for these cross sections.

In conclusion, FEM calculations of the gate capacitance of back-gated
NW-FETs using SiO$_2$ are discussed. Compared with 
the analytical model of completely embedded 
NWs, the capacitances of non-embedded NWs are nearly two
times smaller. This can be accounted for by an effective
dielectric constant of 2.20 within the analytical model of embedded
NWs. However, for high-$k$ materials as HfO$_2$ this is not
possible and the analytical solution cannot be applied. 
Due to the electric field amplification of non-embedded NWs, the
dielectric strength of SiO$_2$ is reached locally at much lower gate
voltages than for the planar geometry. Thus the gate control of
the charge density inside the NW is drastically improved by embedding
the NW. Furthermore, non-embedded NWs
with non-circular cross sections have been considered.

I acknowledge very fruitful discussions with the members of the
nanowire project team at the Philips Research Laboratories.

\newpage


\begin{thebibliography}{99}
\bibitem{FEM}
Finite element method calculations are performed using COMSOL
Multiphysics v3.2. The capacitance is calculated via an
integration of the electric field energy density $W_e$
($C=2W_e/U_{gs}^2$) as well as via an
integration of the induced charge density $Q$ on both electrodes 
($C=Q/U_{gs}$).
\end{thebibliography}
\end{document}